# Figame: A Family Digital Game Based on JME for Shaping Parent-Child Healthy Gaming Relationship

Liyi Zhang[1, *], Yujie Peng[1], Yi Lian[2], Mengru Xue[1]

1 Zhejiang University Ningbo Innovation Center, China
2 NingboTech University, China

With the development of technology, digital games have permeated into family and parent-child relationships, leading to cognitive deficiencies and inter-generational conflicts that have yet to be effectively addressed. Building on previous research on digital games and parent-child relationships, we have developed Figame, a Joint Media Engagement (JME) based parent-child digital game aimed at fostering healthy family gaming relationships through co-playing experiences. The game itself involves providing game-related cognitive support, facilitating role-switching between parent and child, encouraging discussions both within and outside the game, and balancing competition and collaboration. During the study, we assessed the gameplay experiences of 8 parent-child pairs (aged between 8 and 12 years). The results indicated that Figame effectively enhances parent-child digital gaming relationships and promotes a willingness to engage in shared gameplay, thereby fostering positive family dynamics within the context of digital gaming.

Keywords: Parent-Child Interaction, Joint Media Engagement, Game Design

## Introduction

In the era of technological advancements, various forms of entertainment have impacted parent-child relationships differently. Digital games have become an integral part of the lives of many children, particularly pre-adolescents between the ages of 8 and 12 who consume a significant amount of their time on it [13]. Parent-child interactions require more effort and attention in the context of discussing and engaging with digital games [32,35]. Additionally, family dynamics and parental influence often play a significant role in shaping children's gaming habits, and positive family gaming relationships can promote parent-child relationships [18].

Based on previous research findings, it has been observed that Joint Media Engagement (JME) within families tends to promote healthy family relationships on various levels. Some applications have attempted to incorporate JME approaches into educational apps to support children's learning within the family [37,38]. However, they often overlook the existing issues related to digital gaming within families and the untapped potential of JME in problem-solving.

We conducted a field study with 8 parent-child pairs at a community center to explore how game design affects family dynamics during Figame gameplay. Our research translates JME strategies into practical gaming methods and observes their impact in real-world settings.

## Related Work

Digital games, encompassing computer, video, mobile, and online games, often provide multiplayer modes that emphasize player interaction [39,5]. Family environments significantly influence parent-child gaming habits, with poor relationships and authoritarian parenting increasing the risk of gaming addiction in children [2]. Research typically focuses on children aged 7 to 16, whose cognitive development is still maturing [35,39], highlighting the need for family-inclusive solutions to address digital gaming issues [40].

Over the years, studies have explored various family engagements with media, including communication and play, focusing on psychological, social, and learning outcomes [3,5,7,10,15,20,34,40,41]. Recent findings suggest digital games can positively affect family relationships and reduce children's social anxiety, promoting positive family interactions [1,11,29,33,42]. However, executing family gaming is challenging, and no comprehensive methodology has been widely applied [43]. This gap motivates the exploration of specific applied methodologies for parent-child co-play.

Recent research has shifted towards diverse parent-child interaction patterns beyond parental guidance [8]. Joint Media Engagement (JME) involves simultaneous media use by multiple individuals, fostering communication and bonding [26,23,28]. JME facilitates dialogue and relationships between parents and children [12]. Studies have identified key patterns for healthy gaming relationships, such as intimate gaming spaces and balanced cooperation/competition [16,24]. This paper proposes developing parent-child games using JME strategies, emphasizing shared technology's role in modern family bonds to enhance parent-child connections through gaming.

## FAMILY DIGITAL GAME BASED ON JME

This paper explores the integration of JME into digital games to enhance parent-child relationships. We developed "Figame," a 2D game designed to promote gaming habits and cooperative interaction, based on previous research [24]. The game, created using the Unity engine, features mutual rule introductions, role-switching, and elements of competition and cooperation to transform power dynamics and enhance relationship perceptions.

The gameplay involves a parent and child controlling characters to eliminate grid squares and open treasure chests within a time limit, with the goal of earning points and completing tasks. The game consists of two rounds, with a five-minute discussion after each round. The first discussion focuses on game content, while the second addresses gaming habits in daily life. The game concludes with a tally of points to determine the winner.

Our research identified a significant gap in digital game content cognition between parents and children, with children often having a richer understanding due to their interest, while parents face cognitive challenges. To bridge this gap, Figame uses JME by having one party explain game rules to the other within a limited time, fostering interactive involvement and reducing cognitive barriers.

Traditionally, parent-child gaming roles are stable, with parents leading; however, dynamic JME relationships involve flexible roles that enhance mutual understanding. Figame incorporates tasks that require alternating guidance between parents and children, promoting leadership role shifts and a balanced gaming experience. Additionally, balancing competition and cooperation is crucial. Figame addresses this by having players work towards a shared goal of unlocking treasure chests and collecting stars within a time limit to prevent failure, while also earning individual points for eliminating grid cells, fostering cooperative efforts and competitive achievements.

JME also emphasizes facilitating communication and shared experiences. In Figame, post-round discussions focus on game content and are followed by a final discussion about gaming habits in family life, encouraging attention to the game's content, reasons for playing, and consideration of management strategies for gaming within the family context.

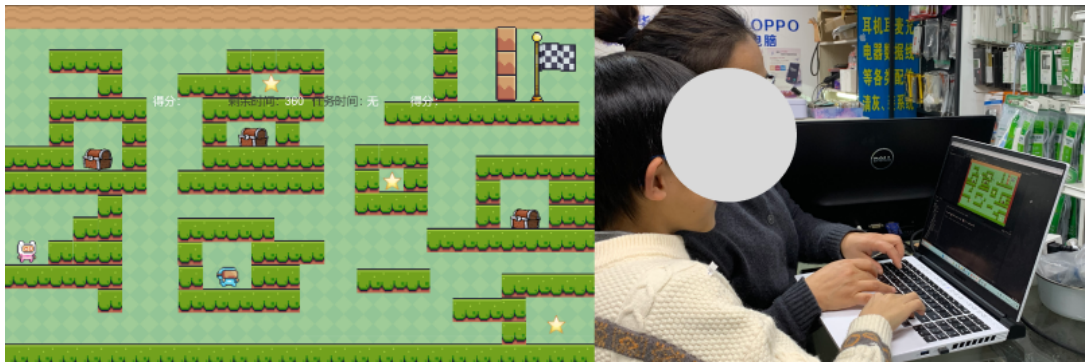

Figure 1: Screenshot of the game(left), The process of parent-child participation in games(right).

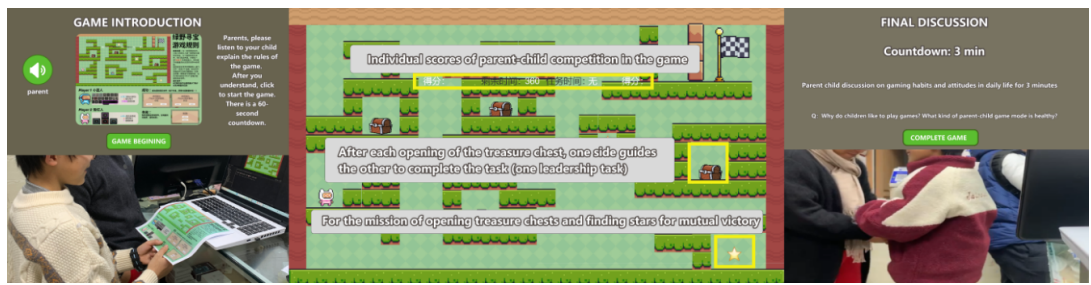

Figure 2: Parent-child interaction where one party explains to the other to understand the rules(left), Figame simultaneously incorporates competition and cooperation and switches the leadership between parents and children during the game (middle), A 3-minute rule explanation time is provided for both during and after the game to enhance parent-child communication(right).

## Method

We recruited 8 parent-child pairs through community centers, targeting families with established gaming habits. The children's ages ranged from 8 to 12 years old.

Our research aims to promote healthy parent-child interaction and gaming habits through parental involvement in games. We conducted an experiment using Figame to evaluate its impact on these interactions and its effectiveness in fostering healthy gaming habits.

In the Figame experiment, parent-child pairs were introduced to the game's background and rules, then engaged in cooperative and competitive activities to complete tasks. Participants had three minutes to understand random rules before explaining them to their partner, aiming to improve mutual understanding. This was followed by two game rounds, with a three-minute discussion after each to share experiences and strategies. The final discussion focused on the broader impact of gaming on their relationship and real-life applications of in-game strategies.

After the game and discussions, participants completed a Likert scale questionnaire to rate the effectiveness of Figame in shaping healthy gaming habits. The questionnaire, based on their experiences and observations, assessed the game's impact on parent-child relationships and gaming habits.

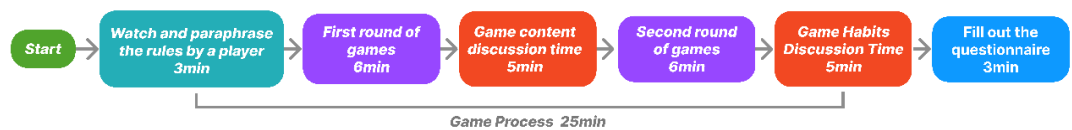

Figure 3: Figame testing process flowchart

# Finding

In our experiment, interactive explanations of game rules significantly boosted engagement and focus for parent-child pairs, making the game content clearer and enhancing the parent-child bond. Most pairs, like P2, P5, and P8, showed increased communication and patience, leading to better interaction and cooperation. P6's father remarked, "This is the first time I understand the rules of the game and play smoothly with my son," highlighting the benefits of clear game explanations.

In Figame, children often lead their parents through special tasks, showcasing their abilities and strengthening the parent-child bond. A parent of P4 was surprised by their child's leadership in task completion, and a child from P1 noted a different feeling when playing with parents, enhancing engagement and excitement. This role reversal fosters interaction, cooperation, and mutual understanding.

Parents and children also showed higher tolerance for game-related errors. P6's parent noted that solving game issues together felt collaborative rather than blameful, turning frustrations into enjoyable challenges. P2's child found humor in game glitches when playing with their father, reflecting a positive attitude towards imperfections.

Game content served as a catalyst for enhanced dialogue about gaming habits and strategies. P6 and P7 improved their performance and comfort in gaming after integrating game-related discussions into daily life. P2's parent realized the importance of companionship over pressure, deciding to communicate more and participate in their child's activities, fostering trust and coordination.

## Discussion and Conclusion

Previous research has demonstrated the effectiveness of Joint Media Engagement (JME) in enhancing parent-child relationships by improving communication and mutual enjoyment. However, most studies have focused on surveys and observational practices rather than practical applications in game design. This study applied JME strategies in game design to improve parent-child relationships, employing communication-oriented rule explanations, role and leadership transitions, balanced cooperation and competition, and enhanced communication inside and outside the game.

The study confirmed that communication-oriented rule explanations help parents better understand digital games, while joint participation in JME reveals fixed daily roles, enhancing children's sense of participation. In Figame, children gain partial leadership, fostering better communication and parental recognition. Balancing competition and cooperation is crucial, as healthy family game relationships require active conflict regulation. Figame facilitated harmony and excitement, encouraging joint play and promoting mutual understanding through in-and-out game communication.

The study had limitations, including a small sample size and unclear indicators for identifying which parent-child pairs benefit most from Figame, as well as a lack of exploration of the long-term effects on parent-child relationships. This study aimed to bridge the intergenerational gap in family contexts through the JME-based game, Figame. Evaluated with eight parent-child pairs, Figame effectively enhanced parent-child digital game relationships and encouraged joint gameplay, fostering healthy family dynamics. The findings offer design insights for future research on parent-child digital games.